\documentclass[prd,nofootinbib,twocolumn]{revtex4}
\usepackage{amsmath}

\begin{document}

\title{Self-interaction spin effects in inspiralling compact binaries}
\author{Bal\'{a}zs Mik\'{o}czi$^{1}$, M\'{a}ty\'{a}s Vas\'{u}th$^{2}$, L\'{a}%
szl\'{o} \'{A}. Gergely$^{1}$}
\affiliation{$^{1}$Departments of Theoretical and Experimental Physics, University of
Szeged, Szeged 6720, Hungary}
\affiliation{$^{2}$KFKI Research Institute for Particle and Nuclear Physics, Budapest
114, P.O.Box 49, H-1525 Hungary}

\begin{abstract}
Gravitational radiation drives compact binaries through an inspiral phase
towards a final coalescence. For binaries with\textit{\ spin, mass
quadrupole and magnetic dipole moments}, various contributions add to this
process, which is characterized by the rate of increase $df/dt$ of the
gravitational wave frequency and the accumulated number $\mathcal{N}$ of
gravitational wave cycles. We present here all contributions to $df/dt$ and $%
\mathcal{N}$\ up to the second post-Newtonian order. Among them we give for
the first time the contributions due to the \textit{self-interaction of
individual spins}. These are shown to be commensurable with the proper
spin-spin contributions for the recently discovered J0737-3039 double
pulsar, and argued to represent the first corrections to the radiation
reaction in the Lense-Thirring approach.
\end{abstract}

\date{\today }
\maketitle
\startpage{1}

\section{Introduction}

Neutron-stars and black-holes forming compact binary systems emit
gravitational radiation. The frequency range of this radiation is expected
to be in the sensitivity range of the Earth-based interferometric detectors
LIGO \cite{LIGO}, VIRGO \cite{VIRGO}, GEO \cite{GEO}, TAMA \cite{TAMA} and
currently observation is under way to capture such signals. Recently, a
method of setting upper limits on inspiral event rates for binary neutron
stars using interferometer data were established from the first scientific
run of LIGO \cite{LIGONS} and\ limits on gravitational wave emission from
selected pulsars using LIGO data were published \cite{LIGOpulsars}. Also
analysis methods were specified for inspiral signals from binaries with 3-20
solar masses \cite{LIGOBH}. Gravitational radiation from these compact
binaries also is expected to be detected by the Laser Interferometer Space
Antenna (LISA) \cite{LISA}, \cite{LISA1}. Parameters of the spinning compact
binaries can be estimated and alternative theories of gravity can be tested %
\cite{BBW} from these measurements.

The final cataclysmic coalescence of such compact binaries is preceded by a
milder inspiral phase, for which the post-Newtonian (PN) approach provides a
reliable description. This description is generally considered valid until
the system reaches the innermost stable circular orbit (ISCO). For neutron
star binaries with realistic equation of state the gravitational wave
frequency at ISCO is between $800-1230$ Hz, according to \cite%
{Gourgoulhon1,Gourgoulhon2,Oechslin}. The upper limit of senstivity for the
LIGO\ detector is $\sim1000$ Hz. The simple estimate shows that for
a neutron star of $1.4$ solar masses the ISCO (at three Schwarzschild radii)
is of about $2.5$ times the characteristic radius of $10$ km of the neutron
star. A more sophisticated argument, taking into account various equations
of state for the neutron stars \cite{Berti} shows $r_{ISCO}/r_{NS}\in
(1.1,2.2)$. Buonanno, Chen and Vallisneri have discussed the problem of the
failure of the PN expansion during the last stages of inspiral (the
intermediate binary black hole, IBBH problem \cite{IBBH}) both for
non-spinning and for spinning black hole binaries on quasicircular orbits %
\cite{BCV1,BCV2}. Reliable results can be achieved by a frequency cutoff,
e.g. stopping the integration at the minimum of the energy as function of
orbital frequency, at ICO \cite{Blanchet02} (called MECO in \cite{BCV2}). In
even latter stages of the inspiral, tidal torques become important \cite%
{Kochanek,BC}. Then each component can be represented by a Schwarzschild
metric corrected with the influence of the Weyl curvature generated by the
other component \cite{Detweiler,Poisson04}. The corrected metric was
computed to second order in the inverse of the Weyl curvature radius in \cite%
{Alvi,Yunes} and to third order by Poisson \cite{Poisson05}.

In the post-Newtonian regime, the equations of motion were given to 3.5 PN
order accuracy, with the inclusion of spin-orbit (SO) effects and their
first PN correction in \cite{Will}. Spin-spin (SS) \cite{Kidder},
quadrupole-monopole (QM) \cite{Poisson} and magnetic dipole - magnetic
dipole (DD) contribution \cite{IT} to the accelerations were also discussed.
The backreaction on the orbit of the gravitational waves escaping compact
binaries can be characterized by the rate of radiative change of otherwise
conserved quantities, like the \textit{secular} energy loss $\langle
dE/dt\rangle $ and \textit{secular} angular momentum loss $\langle d\mathbf{J%
}/dt\rangle $. These were computed over the years with great accuracy. The
leading order contribution was given by \cite{Peters}, \cite{PM}, its first
post-Newtonian (PN) correction in \cite{DD}, while the second post-Newtonian
(2PN) correction in \cite{GI} (a correction is expected, see \cite{Gopa}).

Taking into account the interaction of the spins $\mathbf{S}_{\mathbf{i}}$
with the orbit, the total angular momentum $\mathbf{J}=\mathbf{L}+\mathbf{S}%
_{\mathbf{1}}+\mathbf{S}_{\mathbf{2}}$ is still conserved, however the
orbital angular momentum $\mathbf{L}$ is not, due to spin precessions.
Still, its magnitude $L$ remains a constant of motion, as consequence of the
specific functional form of the precession equation \cite{GPV3}. The
radiative changes $\langle dE/dt\rangle $ and $\langle dL/dt\rangle $
characterize the backreaction on the \textit{radial} part of the motion.
They were computed in \cite{GPV3} and \cite{RS} (these approaches rely on
different spin supplementary conditions and distinct averaging techniques
over a radial orbit - the results, however agree). The secular losses $%
\langle dE/dt\rangle $ and $\langle dL/dt\rangle $ depend on the relative
angles $\kappa _{i}$ of the spins and orbital angular momentum. In order to
have a closed system of differential equations their radiative evolution
(together with the radiative evolution of the angle $\gamma $, subtended by
the two spins) has been given in \cite{GPV3}. For compact binaries the SO
contribution occurs at $\epsilon ^{3/2}$ order, $\epsilon $ representing one
post-Newtonian order.

The SS contribution to $\langle dE/dt\rangle $ and $\langle dL/dt\rangle $
and the subsequent evolution of the angles $\kappa _{i}$ and $\gamma $ were
derived in \cite{spinspin1}, \cite{spinspin2}. Here a description in terms
of the magnitude of an (angular averaged) orbital angular momentum $\bar{L}$
was introduced, this convenient method being kept in the discussion of the
QM contribution \cite{quadrup} and the DD contribution \cite{mdipole}. (The
magnitude $L$ of the orbital angular momentum is not conserved due to the
spin precessions caused by spin-spin, quadrupole-monopole and magnetic
dipole - magnetic dipole interactions.) The SS, QM and DD contributions all
arise at $\epsilon ^{2}$\ order.

Remarkably, the SS contributions in $\langle dE/dt\rangle $ and $\langle
dL/dt\rangle $ given in \cite{spinspin1} contained not only \textit{%
interaction terms between the two spins}, but \textit{self-interaction spin
terms} (SS-self) as well. These arose from the terms proportional to $%
J_{SO}^{\left( 3\right) jl}\left( \mathbf{a}_{N}\right) J_{SO}^{\left(
3\right) jl}\left( \mathbf{a}_{N}\right) $ in $dE/dt$ and $\epsilon
^{ijk}J_{SO}^{\left( 2\right) jl}\left( \mathbf{a}_{N}\right) J_{SO}^{\left(
3\right) kl}\left( \mathbf{a}_{N}\right) $ in $d\mathbf{J}^{i}/dt$ (here $%
J_{SO}^{\left( n\right) jl}$ denotes the $n^{th}$ derivative of the
spin-orbit contribution of the velocity quadrupole moment evaluated with the
Newtonian acceleration $\mathbf{a}_{N}$; there is summation understood over
the repeated indices $j,k,l$). These terms are of order $\epsilon ^{2}$ as
well \cite{Kidder1}.

Although the SS-self contributions to the closed system of differential
equations $\langle d\left( E,L,\kappa _{i},\gamma \right) /dt\rangle $ has
been given in \cite{spinspin1}, \cite{spinspin2}, these terms are
notoriously missing from the bookkeeping of different contributions to
various results concerning inspiralling compact binaries. In this paper we
compute the SS-self contributions to the rate of increase of the
gravitational wave frequency $f$ and to the accumulated number of
gravitational wave cycles $\mathcal{N}$. For completeness we also enlist all
other contributions to $df/dt$ and $\mathcal{N}$ to order $\epsilon ^{2}$
(the PN, SO, SS, QM, DD, 2PN and tail contributions).

On the long run due to the emission of gravitational waves the orbit tends
to circularize \cite{Peters}. Therefore we consider quasicircular orbits,
for which the gravitational wave frequency is twice the orbital frequency%
\cite{Thorne}. In Sec.II first we evaluate the rate of increase of $f$. This
is given by the rate of change of the orbital angular frequency $\omega =\pi
f$ under radiation reaction:%
\begin{equation}
\left( \frac{d\omega }{dt}\right) ^{circ}=\left( \frac{dE}{d\omega }\right)
^{-1}\left\langle \frac{dE}{dt}\right\rangle ^{circ}\ ,  \label{omegadot}
\end{equation}%
where the expression $dE/d\omega $ can be found by differentiating $%
E=E\left( \omega \right) $. We have verified that the circular orbit limit
of the \textit{instantaneous} energy loss $\left( dE/dt\right) ^{circ}$ is
the same as the circular orbit limit of the \textit{secular} energy loss $%
\left\langle dE/dt\right\rangle ^{circ}$, the latter however is much simpler
to compute. Here the various contributions to the secular energy loss 
\begin{eqnarray}
\left\langle \frac{dE}{dt}\right\rangle  &=&\left\langle \frac{dE}{dt}%
\right\rangle _{N}+\left\langle \frac{dE}{dt}\right\rangle
_{PN}+\left\langle \frac{dE}{dt}\right\rangle _{SO+tail}  \notag \\
&&+\left\langle \frac{dE}{dt}\right\rangle
_{2PN+(SS-self)+S_{1}S_{2}+QM+DD}\   \label{Eloss}
\end{eqnarray}%
were given before in \cite{DD}, \cite{GI}, \cite{GPV3}-\cite{mdipole} and %
\cite{BlanchetSchafer}. We have grouped together the contributions occuring
at the same post-Newtonian order.

Eq. (\ref{omegadot}) being an ordinary differential equation in $\omega $
can be immediately integrated. Then we compute the accumulated number of
gravitational wave cycles $\mathcal{N}$ by a second integration. Finally we
evaluate these contributions to $\mathcal{N}$ for some famous or typical
compact binary systems discussed in the literature, and we compare the
magnitude of the SS-self contribution with other contributions. This is done
in the Concluding Remarks.

In the following we define some of the notations employed in the paper.

The magnitude and direction of $\mathbf{S}_{\mathbf{i}}$ are denoted as $%
S_{i}$ and $\mathbf{\hat{S}}_{\mathbf{i}}$. Then $\gamma =\cos ^{-1}(\mathbf{%
\hat{S}_{1}\cdot \hat{S}_{2})}$. The magnitude and direction of the magnetic
dipole moments are denoted as $d_{i}$ and $\mathbf{\hat{d}_{i}}$. They
subtend the angle $\lambda =\cos ^{-1}(\mathbf{\hat{d}_{1}\cdot \hat{d}_{2})}
$ with each other.\ In a coordinate systems $\mathcal{K}$ with the axes $(%
\mathbf{\hat{c},\hat{L}\times \hat{c},\hat{L}})$, where $\mathbf{\hat{c}}$
is the unit vector in the $\mathbf{J\times L}$ direction, the polar angles $%
\kappa _{i}$ and $\psi _{i}$ of the spins are defined as $\mathbf{\hat{S}%
_{i}=}(\sin \kappa _{i}\cos \psi _{i},\sin \kappa _{i}\sin \psi _{i},\cos
\kappa _{i})$. We also have introduced in \cite{mdipole} coordinate systems $%
\mathcal{K}^{i}$ with the axes $(\mathbf{\hat{b}_{i},\hat{S}_{i}\times \hat{b%
}_{i},\hat{S}_{i}})$, where $\mathbf{\hat{b}_{i}}$ are the unit vectors in
the $\mathbf{S_{i}\times L}$ directions, respectively. The polar angles $%
\alpha _{i}$ and $\beta _{i}$ of the the magnetic dipole moments $\mathbf{{d}%
_{i}\,}$in the coordinate systems $\mathcal{K}^{i}$ are $\mathbf{{\hat{d}}%
_{i}=}(\sin \alpha _{i}\cos \beta _{i},\sin \alpha _{i}\sin \beta _{i},\cos
\alpha _{i})$. Quadrupolar contributions are characterized by the parameters 
$p_{i}=Q_{i}/m_{i}m^{2}$, where $Q_{i}$ is the quadrupole-moment scalar \cite%
{Poisson} of the $\ i^{th}$ axially symmetric binary component with symmetry
axis $\mathbf{\hat{S}}_{\mathbf{i}}$. (In the Newtonian limit $Q_{i}=\Theta
_{i}^{\prime }-\Theta _{i}=-S_{i}\left( \Theta _{i}/\Theta _{i}^{\prime
}-1\right) /\Omega _{i}$ for each axisymmetric body characterized by the
principal moments of inertia $\left( \Theta _{i},\ \Theta _{i},\ \Theta
_{i}^{\prime }\right) $ and angular velocity $\Omega _{i}=S_{i}/\Theta
_{i}^{\prime }$.)

Throughout the paper we use $G=c=1$.

\section{Frequency evolution and the accumulated number of cycles}

For circular orbits $\dot{r}=\ddot{r}=0$ and $v^{3}=m\omega $ holds.
Consequently $\left( m\omega \right) ^{2/3}$ is of order $\epsilon $. The
radial projection of the acceleration defines the \textit{orbital angular
velocity} as $\mathbf{r\cdot a}=-r^{2}\omega ^{2}$ \cite{Kidder}. From the
explicit form of the acceleration (with various contributions given in \cite%
{Kidder}, \cite{Poisson} and \cite{IT}) we then find $\omega =\omega \left(
r\right) $ and subsequently $r=r\left( \omega \right) $ as 
\begin{eqnarray}
r(\omega ) &=&m(m\omega )^{-2/3}\Biggl\{1-\frac{3-\eta }{3}(m\omega )^{2/3} 
\notag \\
&&-\frac{m\omega }{3}\sum_{i=1}^{2}\left( 2\frac{m_{i}^{2}}{m^{2}}+3\eta
\right) \frac{S_{i}}{m_{i}^{2}}\cos \kappa _{i}  \notag \\
&&-(m\omega )^{4/3}\Bigg[-\eta \left( \frac{19}{4}+\frac{\eta }{9}\right) 
\notag \\
&&+\frac{S_{1}S_{2}}{2\eta m^{4}}\left( \cos \gamma -3\cos \kappa _{1}\cos
\kappa _{2}\right)  \notag \\
&&+\frac{1}{4}\sum_{i=1}^{2}p_{i}\left( 3\cos ^{2}\kappa _{i}-1\right) +%
\frac{d_{1}d_{2}\mathcal{A}_{0}}{2\eta m^{4}}\Biggr]\Biggr\}\ .
\label{romega}
\end{eqnarray}%
In the magnetic dipolar contribution we have introduced the shorthand
notation:%
\begin{eqnarray}
\mathcal{A}_{0} &=&2\cos \lambda +3(\rho _{1}\sigma _{2}-\rho _{2}\sigma
_{1})\sin (\psi _{2}-\psi _{1})  \notag \\
&&-3(\rho _{1}\rho _{2}+\sigma _{1}\sigma _{2})\cos (\psi _{2}-\psi _{1})\ ,
\label{Bk}
\end{eqnarray}%
with 
\begin{eqnarray}
\rho _{i} &=&\sin \alpha _{i}\cos \beta _{i}\ ,  \notag \\
\sigma _{i} &=&\cos \alpha _{i}\sin \kappa _{i}+\sin \alpha _{i}\sin \beta
_{i}\cos \kappa _{i}\ .
\end{eqnarray}%
Combining Eq. (\ref{romega}) with the expression of the velocity on circular
orbits $v=r\omega $ in the energy integral $E=E\left( r,v\right) $ we obtain 
$E=E\left( \omega \right) $:%
\begin{eqnarray}
E(\omega ) &=&-{\frac{1}{2}}\mu (m\omega )^{2/3}\Biggl\{1-\frac{1}{4}\left(
3+\frac{\eta }{3}\right) (m\omega )^{2/3}  \notag \\
&&+m\omega \sum_{i=1}^{2}\left( \frac{8}{3}\frac{m_{i}^{2}}{m^{2}}+2\eta
\right) \frac{S_{i}}{m_{i}^{2}}\cos \kappa _{i}  \notag \\
&&+(m\omega )^{4/3}\Biggl[{\frac{1}{8}}(-27+19\eta -\frac{\eta ^{2}}{3}) 
\notag \\
&&+\frac{S_{1}S_{2}}{\eta m^{4}}\left( \cos \gamma -3\cos \kappa _{1}\cos
\kappa _{2}\right)  \notag \\
&&+\frac{1}{2}\,\sum_{i=1}^{2}p_{i}(3\cos ^{2}\kappa _{i}-1)+\frac{d_{1}d_{2}%
\mathcal{A}_{0}}{\eta m^{4}}\Biggr]\Biggr\}\ .  \label{Eomega}
\end{eqnarray}%
The radiative orbital angular frequency evolution for the above defined
circular orbits is deduced from Eqs. (\ref{omegadot}), (\ref{Eloss}) and (%
\ref{Eomega}), with the various contributions given as 
\begin{eqnarray}
\left\langle \frac{d\omega }{dt}\right\rangle ^{circ} &=&\frac{96\eta
m^{5/3}\omega ^{11/3}}{5}\Biggl[1-\left( \frac{743}{336}+\frac{11}{4}\eta
\right) \left( m\omega \right) ^{2/3}  \notag \\
&&+\left( 4\pi -\beta \right) m\omega +\Biggl(\frac{34103}{18144}+\frac{13661%
}{2016}\,\eta  \notag \\
&&+\frac{59}{18}\,\eta ^{2}+\sigma \Biggr)\left( m\omega \right) ^{4/3}%
\Biggr]\ ,  \label{dotomega}
\end{eqnarray}%
where%
\begin{equation}
\sigma =\sigma _{S_{1}S_{2}}+\sigma _{SS-self}+\sigma _{QM}+\sigma _{DD}\ .
\end{equation}%
The quantities $\beta $, $\sigma _{S_{1}S_{2}}$, $\sigma _{SS-self}$, $%
\sigma _{QM}$ and $\sigma _{DD}$ are the spin-orbit, spin-spin,
self-interaction spin, quadrupole-monopole and magnetic dipole-dipole
parameters, respectively: 
\begin{subequations}
\begin{gather}
\beta =\frac{1}{12}\sum_{i=1}^{2}\frac{S_{i}}{m_{i}^{2}}\left( 113\frac{%
m_{i}^{2}}{m^{2}}+75\eta \right) \cos \kappa _{i}\ ,  \label{beta} \\
\sigma _{S_{1}S_{2}}=\frac{S_{1}S_{2}}{48\eta m^{4}}(-247\cos \gamma
+721\cos \kappa _{1}\cos \kappa _{2})\ ,  \label{SS} \\
\sigma _{SS-self}=\frac{1}{96m^{2}}\sum_{i=1}^{2}\left( \frac{S_{i}}{m_{i}}%
\right) ^{2}\left( 6+\sin ^{2}\kappa _{i}\right) \ ,  \label{self} \\
\sigma _{QM}=-\frac{5}{2}\sum_{i=1}^{2}p_{i}\left( 3\cos ^{2}\kappa
_{i}-1\right) \ ,  \label{QM} \\
\sigma _{DD}=-\frac{5}{\eta m^{4}}d_{1}d_{2}\mathcal{A}_{0}\ ,  \label{siDD}
\end{gather}%
The N, PN, SO, SS, 2PN and tail contributions to Eq. (\ref{dotomega}) were
verified to agree with those given in \cite{PoissonWill}, the QM with those
in \cite{Poisson} and the DD with those in \cite{IT}, respectively. We
emphasize the \textit{SS-self contribution}, given for the first time.

Eq. (\ref{dotomega}) is an ordinary differential equation in $\omega $,
provided all angular variables can be considered constants. To linear order
in the perturbations this is true. This is because all angles appear only in
the perturbative corrections, therefore they are given with sufficient
accuracy to $\epsilon $ order, in which order they are constants \cite{GPV3}%
. Thus the time-evolution of the orbital angular frequency $\omega $ for
circular orbits can be easily deduced by an integration over time: 
\end{subequations}
\begin{eqnarray}
\omega (t) &=&\frac{\tau ^{-3/8}}{8m}\Biggl\{1+\left( \frac{743}{2688}+\frac{%
11}{32}\eta \right) \tau ^{-1/4}  \notag \\
&&+\frac{3}{10}\left( \frac{\beta }{4}-\pi \right) \tau ^{-3/8}+\Biggl(\frac{%
1855099}{14450688}+\frac{56975}{258048}\eta  \notag \\
&&+\frac{371}{2048}\eta ^{2}-\frac{3\sigma }{64}\Biggr)\tau ^{-1/2}\Biggr\}\
,  \label{omega}
\end{eqnarray}%
where the dimensionless time variable $\tau =\eta (t_{c}-t)/5m$ is related
to the time $(t_{c}-t)$ left until the final coalescence. A further
integration gives the accumulated orbital phase $\phi _{c}-\phi $:%
\begin{equation}
\phi _{c}-\phi =\frac{5m}{\eta }\int \omega (\tau )d\tau \ .
\end{equation}%
From here the accumulated number of gravitational wave cycles emerges as 
\begin{eqnarray}
\mathcal{N} &=&\frac{\phi _{c}-\phi }{\pi }=\frac{1}{\pi \eta }\Biggl\{\tau
^{5/8}+\left( \frac{3715}{8064}+\frac{55}{96}\eta \right) \tau ^{3/8}  \notag
\\
&&+\frac{3}{4}\left( \frac{\beta }{4}-\pi \right) \tau ^{1/4}+\Biggl(\frac{%
9275495}{14450688}+\frac{284875}{258048}\eta  \notag \\
&&+\frac{1855}{2048}\eta ^{2}-\frac{15\sigma }{64}\Biggr)\ \tau ^{1/8}%
\Biggr\}  \label{N}
\end{eqnarray}%
The tail and 2PN contributions agree with those given in \cite{Blanchet95},
the rest of the terms with Eq. (4.16) of \cite{Kidder}. We also note that
the numerical values in Table II of \cite{BBW} could be recovered from our
Eq. (\ref{N}).

In Table \ref{Table1} we enlist all contributions to $\mathcal{N}$ in terms
of $\beta $ and $\sigma $, evaluated numerically from our Eq. (\ref{N}) for
several compact binaries. These are two well-known binary neutron star
systems, the newly discovered double pulsar J0737-3039 (\cite{pulsar1}, \cite%
{pulsar2}, and for a recent review see \cite{pulsar3}), the Hulse-Taylor
pulsar B1913+16, one neutron star - stellar mass black hole binary, one
binary composed of two stellar mass black holes \cite{BDIWW}, finally two
examples of galactic black hole - galactic black hole binaries \cite{BBW}.

\section{Concluding remarks}

We have presented the complete set of contributions up to $\epsilon ^{2}$
order (PN, SO, SS, QM, DD, tail and 2PN) to the evolution of gravitational
wave frequency and to the accumulated number of gravitational wave cycles,
with the inclusion of the previously unknown self-interaction spin terms.
These results add to the already derived \cite{GPV3}, \cite{spinspin1}-\cite%
{mdipole} closed system of first order differential equations governing the
secular evolution of radiating compact binaries and represent an important
step towards a complete characterization of the orbital evolution.

\begin{widetext}
\begin{table}[h]
\caption{Number of accumulated gravitational wave cycles, either for the frequency 
domain of the detectors or up to the innermost stable circular orbit. The first four sets 
of data refer to the LIGO/VIRGO type detectors, while the last two columns to the 
LISA bandwidth.}
\label{Table1}%
\begin{equation*}
\begin{tabular}{|c|c|c|c|c|c|c|}
\hline
PN Order & $%
\begin{array}{c}
\text{J0737-3039} \\ 
1.337M_{\odot } \\ 
1.250M_{\odot }%
\end{array}%
$ & $%
\begin{array}{c}
\text{B1913+16\smallskip\ } \\ 
1.387M_{\odot } \\ 
1.441M_{\odot }%
\end{array}%
$ & $%
\begin{array}{c}
\text{BDIWW1} \\ 
1.4M_{\odot } \\ 
10M_{\odot }%
\end{array}%
$ & $%
\begin{array}{c}
\text{BDIWW2} \\ 
10M_{\odot } \\ 
10M_{\odot }%
\end{array}%
$ & $%
\begin{array}{c}
\text{BBW1} \\ 
10^{4}M_{\odot } \\ 
10^{5}M_{\odot }%
\end{array}%
$ & $%
\begin{array}{c}
\text{BBW2} \\ 
10^{7}M_{\odot } \\ 
10^{7}M_{\odot }%
\end{array}%
$ \\ \hline\hline
$f_{in}(Hz)$ & $10$ & $10$ & $10$ & $10$ & $4.199\times 10^{-4}$ & $%
1.073\times 10^{-5}$ \\ \hline
$f_{fin}(Hz)$ & $1000$ & $1000$ & $360$ & $190$ & $3.997\times 10^{-2}$ & $%
2.199\times 10^{-4}$ \\ \hline\hline
$N$ & $18310$ & $15772.1$ & $3580$ & $600$ & $21058$ & $535$ \\ \hline
$PN$ & $475.8$ & $435$ & $212$ & $59$ & $677$ & $55$ \\ \hline
$SO$ & $17.5\beta $ & $16.5\beta $ & $14\beta $ & $4\beta $ & $36\beta $ & $%
4\beta $ \\ \hline
$SS-self,SS,QM,DD$ & $-2.1\sigma $ & $-2.1\sigma $ & $-3\sigma $ & $-\sigma $
& $-5\sigma $ & $-\sigma $ \\ \hline
$Tail$ & $-208$ & $-206$ & $-180$ & $-51$ & $-450$ & $-48$ \\ \hline
$2PN$ & $9.8$ & $9.5$ & $10$ & $4$ & $18$ & $4$ \\ \hline
\end{tabular}%
\end{equation*}%
\end{table}
\end{widetext}

In what follows we would like to comment on the importance of the
self-interaction spin contributions. For this, we evaluate the spin
parameters $\sigma _{S_{1}S_{2}}$ and $\sigma _{SS-self}$ for the case of
the recently discovered double pulsar J0737-3039. The neutron stars in this
double pulsar have average radii $15$ km, masses of $1.337$ $M_{\odot }$ and 
$1.25$ $M_{\odot }$, pulse periods of $22.7$ ms and $2773.5$ ms. According
to the Jenet-Ransom model \cite{JenetRansom} the angle $\kappa _{1}$ (in
their notation $\lambda $)\ has two possible values: $\kappa _{1}=167^{\circ
}\pm 10^{\circ }$ (Solution 1), and $\kappa _{1}=90^{\circ }\pm 10^{\circ }$
(Solution 2).\ The other angle $\kappa _{2}$ in principle can be determined
by solving numerically the spin precession equations. This was done for
black hole binaries in\ \cite{schnittman}. However according to \cite%
{pulsar4} and \cite{pulsar5} it is likely that wind-torques from the
energetically dominant component have driven the spin axis of the other
component to align with the direction of $\mathbf{L}$, causing $\kappa
_{2}=0 $. Therefore $\gamma =\kappa _{1}$. We give the estimates for the
spin parameters in Table \ref{Table2}. 
\begin{table}[h]
\caption{The spin parameters for the two solutions of the Jenet-Ransom model
representing the binary pulsar J0737-3039.}
\label{Table2}%
\begin{equation*}
\begin{tabular}{|l|l|l|}
\hline
Spin parameter (order) & Solution 1 & Solution 2 \\ \hline
$\beta \;$ & $-0.166$ & $0.001$ \\ \hline
$\sigma _{S_{1}S_{2}}\;(10^{-4})$ & $-0.372$ & $0$ \\ \hline
$\sigma _{SS-self}$\/$\;(10^{-4})$ & $0.298$ & $0.345$ \\ \hline
\end{tabular}%
\end{equation*}%
\end{table}

Two remarkable conclusions stem out from Table \ref{Table2}. First, let us
consider Solution 1. Though four orders of magnitude smaller than the
spin-orbit effects, already considered in \cite{O'Connell}, self-interaction
spin contributions are found to be comparable with the proper spin-spin
contributions. This happens in spite of the fact that the numerical factors
in $\sigma _{SS-self}$ are two orders of magnitude smaller than those in $%
\sigma _{S_{1}S_{2}}$, cf. Eqs. (\ref{SS}) and (\ref{self}). The
commensurability of the proper and self SS contributions is due to the fact
that one of the spins is two orders of magnitude larger than the other.

Second, consider Solution 2, for which Eq. (\ref{SS}) gives $\sigma
_{S_{1}S_{2}}=0$. (We have checked that even at the limits $\kappa
_{1}=80^{\circ }$ and $100^{\circ }$ of the allowed domain in Solution 2,
the proper spin-spin contribution is still one order of magnitude smaller
than the self-interaction contribution.) We conclude, that whenever one of
the spins is aligned with the orbital angular momentum and the other spin is
perpendicular to it, the proper spin-spin contribution vanishes. This
happens to be the case for Solution 2 of the double pulsar J0737-3039. Thus $%
\sigma _{SS-self}$ is the only spin-spin contribution in this case. We also
remark that for Solution 2 the self-interaction spin contribution is only
two orders of magnitude smaller than the spin-orbit contribution.\ This is
because only the smaller spin contributes to $\beta $.

The importance of the self-interaction spin parameter is in gravitational
wave astronomy, rather than in the traditional electromagnetic observations.
In \cite{BBW} the phasing function $\Psi \left( f\right) $ was derived up to
2PN order. The coefficient $\sigma $ there denotes the proper spin-spin
parameter. However as demonstrated here, the additional self-spin
contributions given by $\sigma _{SS-self}$ is of the same order of
magnitude, thus it should be also included (together with $\sigma _{QM}$ and 
$\sigma _{DD}$, for high mass quadrupole and magnetic dipole moments,
respectively).

The self spin-spin contributions derived in this paper are more important
whenever one of the spins is negligible compared the other. To leading order
in the (dominant) spin and when the mass of the high-spin component also
dominates over its companion's, this is the Lense-Thirring approximation.
Its first correction in the gravitational radiation is represented exactly
by the SS-self contribution of the higher spin.

\section{Acknowledgements}

This work was supported by OTKA grants no. T046939, F049429 and TS044665.
B.M. wishes to thank the organizers of the VI$^{th}$ SIGRAV Graduate School:
Probing Gravity Theories in Binary Systems, for financial support.

\end{document}